\documentclass{article}

\usepackage{arxiv}

\usepackage[utf8]{inputenc} % allow utf-8 input
\usepackage[T1]{fontenc}    % use 8-bit T1 fonts
\usepackage{hyperref}       % hyperlinks
\usepackage{url}            % simple URL typesetting
\usepackage{booktabs}       % professional-quality tables
\usepackage{amsfonts}       % blackboard math symbols
\usepackage{nicefrac}       % compact symbols for 1/2, etc.
\usepackage{microtype}      % microtypography
\usepackage{lipsum}		% Can be removed after putting your text content
\usepackage{} %for table 
\usepackage{graphicx}
\usepackage{natbib}
\usepackage{doi}
\usepackage{amssymb}
\usepackage{url,hyperref,lineno,microtype,subcaption}
\usepackage[onehalfspacing]{setspace}
\usepackage{graphicx}
\usepackage{CJKutf8}
\usepackage{amsmath}
\usepackage{dirtytalk}
\usepackage{algorithm}
\usepackage{varwidth}
\usepackage{mathtools}
\usepackage{multirow}
\usepackage{siunitx}
\usepackage[utf8]{inputenc}
\usepackage{tabto}
\usepackage{algpseudocode} 
\newcounter{algsubstate}

\title{A Bi-level assessment of Twitter in predicting the results of an election: Delhi Assembly Elections 2020}

\author{Maneet Singh \\
	Department of Computer Science and Engineering \\
	Indian Institute of Technology, Ropar, India \\
	\texttt{2018csz0008@iitrpr.ac.in} \\
	\And S.R.S. Iyengar\\
	Department of Computer Science and Engineering \\
	Indian Institute of Technology, Ropar, India \\
	\texttt{sudarshan@iitrpr.ac.in}
	\And Akrati Saxena\\
	Department of Mathematics and Computer Science \\
	Eindhoven University of Technology, Eindhoven, The Netherlands \\
	\texttt{a.saxena@tue.nl}
	\And Rishemjit Kaur\\
	CSIR-Central Scientific Instruments Organisation, Chandigarh, India \\
	Academy of Scientific and Innovative Research, Ghaziabad, India \\
	\texttt{rishemjit.kaur@csio.res.in} \\
}

\date{}

\renewcommand{\headeright}{}
\renewcommand{\undertitle}{}
\renewcommand{\shorttitle}{\textit{A Bi-level assessment of Twitter in predicting the results of an election} }

\hypersetup{
pdftitle={A template for the arxiv style},
pdfsubject={q-bio.NC, q-bio.QM},
pdfauthor={David S.~Hippocampus, Elias D.~Striatum},
pdfkeywords={First keyword, Second keyword, More},
}

\begin{document}
\maketitle

\begin{abstract}
Elections are the backbone of any democratic country, where voters elect the candidates as their representatives. The emergence of social networking sites has provided a platform for political parties and their candidates to connect with voters in order to spread their political ideas. Our study aims to use Twitter in assessing the outcome of Delhi Assembly elections held in 2020, using a bi-level approach, i.e., concerning political parties and their candidates. We analyze the correlation of election results with the activities of different candidates and parties on Twitter, and the response of voters on them, especially the mentions and sentiment of voters towards a party. The Twitter profiles of the candidates are compared both at the party level as well as the candidate level to evaluate their association with the outcome of the election. We observe that the number of followers and the replies to the tweets of candidates are good indicators for predicting actual election outcome. However, we observe that the number of tweets mentioning a party and the sentiment of voters towards the party shown in tweets are not aligned with the election result. We also use machine learning models on various features such as linguistic, word embeddings and moral dimensions for predicting the election result (win or lose). The random forest model using tweet features provides promising results for predicting if the tweet belongs to a winning or losing candidate.

\end{abstract}

% keywords can be removed


\section{Introduction}

Online social networks are one of the highly popular as well as widely used medium for propagating political agendas in this modern era \cite{carlisle2013social,groshek2017helping}. The success of Barack Obama through extensive political campaigning in the 
US presidential elections held in 2008 and 2012 are some of the most famous examples where the power of social networking websites in winning the elections can be understood \cite{metzgar2009social, kreiss2016seizing}. Even, the adverse publicity on Twitter by Trump in 2016 for winning the elections \cite{ross2020going} highlights the effect of social media on voters, irrespective of the type of campaigning. The study on the 2014 Swedish elections \cite{larsson2017going} compared the use of Twitter and Facebook by the political parties and observed that larger parties achieved relatively more attention on the platform. The general elections held in India in 2019 were also highly impacted by the extensive use of social media \cite{rao_2020}.
%The US presidential election held in 2008 is one of the most famous examples where the power of social media in winning the elections can be understood \cite{metzgar2009social}. Barack Obama defeated John McCain in the elections, and one of the significant differences between the two candidates was that Obama mainly used social media, whereas McCain utilized Television for campaigning. Obama was successful in having a large number of followers or supporters and was able even to collect large donations for campaigning itself. The extensive political campaigning by Obama helped him in connecting with more people, and he again won the presidential elections in 2012 \cite{kreiss2016seizing}. The adverse publicity on Twitter by Trump in 2016 for winning the elections \cite{ross2020going} highlights the effect of social media on voters, irrespective of the type of campaigning. The study on the 2014 Swedish elections \cite{larsson2017going} compared the use of Twitter and Facebook by the political parties and observed that larger parties achieved relatively more attention on the platform. The general elections held in India in 2019 were also highly impacted by the extensive use of social media \cite{rao_2020}.

%As social media is actively used by political parties and their members during the elections, it has enabled the researchers to understand the discussions and their impact on the elections. The past research related to the prediction of election results

The previous work on predicting election outcomes using social media data can be divided into two categories, i.e., party level and candidate level. The first line of research that focuses on predicting results at the party level either analyzes the reaction of voters towards the political parties on social media platforms or studies the behavior of parties on these platforms during the election period. Tumasjan et al. \cite{tumasjan2010predicting} were the first ones to reveal from their study that discussions on Twitter can be used to predict the results of an election. The authors used the tweets related to the German election 2009 and found that the number of tweets mentioning a party is a good indicator of the winner.%, i.e., the party having the highest number of tweets mentioning it will be victorious over other parties participating in the election. The linguistic profiles of the candidates, using the tweets mentioning them, were also studied and were found to be dominated by positive emotions over negative ones. 
Another study \cite{sang2012predicting} claims that using only tweets that mention a single party and retaining only the first tweet of every user will guide in getting closer to the vote share of each party involved in the election. Similarly, the share of positive tweets has also been observed to be a better indicator of election results as compare to the share of negative tweets \cite{bermingham2011using}. The use of machine learning models have also been found to be beneficial in identifying the winning party based on the prediction of sentiments in party-specific tweets \cite{sharma2016prediction}. Besides tweet frequency and sentiment of tweets, the order of the political parties in terms of their seat count has also been predicted using data from Twitter platform \cite{burnap2016140,singh2020can}. %The method of predicting sentiments in the party-specific tweets by machine learning models can be used to identify the winning party \cite{sharma2016prediction}. %Burnap et al. \cite{burnap2016140} analyzed the Twitter data related to the 2015 General elections held in the UK to predict the outcome of the election without having any idea of the actual results. The predicted results may not be precisely accurate, but the authors were able to capture the order of the political parties in terms of their seat count. In the case of the 2017 Punjab assembly elections held in India, the positive sentiment ratio along with the history of election results were able to estimate the seat count distribution among the political parties \cite{singh2020can}. 
The strategy of highlighting various national issues such as unemployment and economy on Twitter by the opposing party might also help in winning the elections \cite{ahmed20162014}.

The second line of research in predicting the outcomes of an election aims at either analyzing the behavior of candidates on social media platforms or assessing the voters' response towards the candidates on these platforms. One of the studies \cite{yang2017equalization} was conducted on the 2010 US midterm elections using the tweets of more than 300 candidates as well as their followers. The engagement of candidates with the voters (or followers) on Twitter in the form of retweets and mentions was found to be linked with the outcome of an election \cite{yang2017equalization}. Bright et al. \cite{bright2020does} analyzed the use of Twitter by around 600 candidates of UK national elections held in 2015 and 2017. The vote percentage of candidates was found to be positively correlated with Twitter usage. According to Franzia \cite{francia2018free}, the surprising victory of Donald Trump during the 2016 presidential election could be associated with his supremacy in the use of free media. The tweets mentioning the candidate names can be used to identify the popularity of the candidate with the help of sentiment analysis as shown in the study analyzing the 2017 French elections \cite{wang2017prediction}. The direct mentions of a candidate name may not be sufficient; instead, using additional mentions through different possible aliases of the names along with election relevant keywords may result in better predictions \cite{gaurav2013leveraging}. The network of users surrounding a candidate on social media through friends and followers can also be relevant for the election outcomes; however, it has very little evidence \cite{cameron2016can}.

There is a lack of research that analyzes the election data at the bi-level, i.e., in terms of both participating political parties and the candidates. Our study aims to analyze the Twitter data and its role in predicting the winners of the election using a bi-level approach. The need for undergoing a bi-level assessment can be understood based on the following two scenarios. First, a candidate could win because of their party or party leaders' popularity and reach among the voters. Second, a candidate may win because of their persona among the voters. For our analysis, we collect and utilize a three-fold dataset related to the Delhi assembly elections held in 2020. Our dataset comprises generic tweets relevant to elections, candidate related tweets, and party related tweets. 
In our paper, we have study the correlation of different parameters, such as tweet mentions, sentiments of voters, and availability and activeness of different Twitter profiles, with the election results.
%In our study, we have assessed parameters like the volume of mentions to a party as well as the kind of sentiment present in the tweets mentioning the parties. Various factors such as availability and activeness on the platform are analyzed both with respect to party as well as candidates. The profile attributes of Twitter handles belonging to political parties, and the candidates are also examined to find their relationship with the winning party and the winning candidates. We have also developed a machine learning model that can predict whether the given tweet is by a winning candidate or not. 
The major contribution of our works are as follows.
%Our study answers the following main questions.
\begin{enumerate}
     \item Verify the applicability of frequently used measures, such as mentions and sentiments on the collected Twitter data, for predicting elections results.
    \item Assess candidate availability and activeness on Twitter and analyze its association with the outcomes of the election.
    \item Compare Twitter Profiles of different political parties and their candidates to highlight the differences and their correlation with the election outcome.
    %\item The analysis show that SVM and RF machine learning models using the textual features of the tweet provide good results for predicting whether the tweet belong to winning or losing candidate, that can be further used to predict the election outcome with respect to a political party. 
    \item Analyze machine learning models on the textual features of the tweet to efficiently predict whether the tweet belong to a winning or a losing candidate, that can be further used to predict the election outcome with respect to a political party. 
    %s to predict whether the given tweet is by a winning candidate or a losing candidate.
    %\item Test machine learning models using the textual features of the tweet to predict whether the tweet belong to a winning or a losing candidate, that can be further used to predict the election outcome with respect to a political party.
    
\end{enumerate}

\section{Dataset}

We collect the Twitter data for the Delhi Assembly elections, 2020 using the Twitter API. We mainly collect three different kinds of datasets pertaining to Delhi Assembly elections as explained below.%\footnote{The dataset will be shared publicly once the paper is accepted}.
%with the help of a streaming API provided by Twitter

\subsection{Election relevant tweets}

The tweets related to assembly elections held in Delhi, capital of India, containing keywords `Delhi' and `Election', were collected. Our collection process resulted in 162,556 tweets spanning from 22nd December 2019 to 7th February 2020, i.e., for a period of 48 days before the election date (8th February 2020). Later, we retained tweets of only those users that belong to Delhi, with the help of geo-location mentioned in their tweets or the city specified explicitly by them in the location field of their profile. The data from the location field has been found to be a valid source in past studies \cite{kulshrestha2012geographic,pavalanathan2015confounds}. There were finally 143,616 tweets in our dataset belonging to Delhi users.

\subsection{Candidate related tweets} 
First, we manually collected the official Twitter handles of the candidates of three major political parties, i.e., BJP (Bhartiya Janta Party), AAP (Aam Aadmi Party), and Congress (Indian National Congress). Since all the candidates were not active on Twitter, we were able to extract 101 Twitter handles out of 203 candidates belonging to all three parties. These Twitter handles were then used to collect all the tweets by the candidates along with the retweets and replies to their tweets, from 21st January 2020 to 7th February 2020 using the streaming services by Twitter. The data collection was started from 21st Jan as all three parties declared their final list of candidates for the elections on this date \cite{news18_2020}. On segregating the tweets related to candidates, there were a total of 14,323 tweets by the candidates themselves along with 928,545 retweets and 323,434 replies to their tweets.
%The purpose of beginning the data collection on the specified date was solely based on the fact that the given date coincides with the date by which all three parties have declared their final list of candidates for the elections \cite{news18_2020}. 

\subsection{Party related tweets} 
The Twitter handles of all three political parties specific to Delhi (@BJP4Delhi, @AAPDelhi, and @INCDelhi) were used to collect all their tweets as well as retweets and replies to their tweets using Twitter Streaming API. In this way, we were able to collect 4,733 tweets by all three parties along with 281,209 retweets  40,893 replies to their tweets. 

% \begin{comment}

% \subsection{Actual Election Results}

% For better understanding of the upcoming results, we mention the actual election outcome of Delhi Assembly Election 2020. 

% \end{comment}

\section{Results and Discussions}

In this section, we will discuss our main findings. For better understanding of the analysis, it is vital to know the actual outcomes of the Delhi Assembly Elections held in 2020. The party that emerged as the winner was AAP (won 62 out of 70 seats), whereas BJP was the second best party (won 8 out of 70 seats), and the Congress did not won any seat in the elections. %the actual outcome of the Delhi Elections held in 2020 is as follows:- AAP emerged the winner, whereas BJP was the second best party. On the other hand Congress did not won any seat in the elections.

\subsection{Political Party Mentions}
%There has been work in the literature \cite{tumasjan2010predicting} suggesting that the number of tweets mentioning a party can help in predicting the results of an election. We tested the applicability of the above observation on our election relevant tweets in the following manner.

We first analyze the impact of the tweets mentioning a party on the election results.

\subsubsection{Using Hashtags and Mentions}

\begin{figure*}
  \centering
  \includegraphics[height=7cm,width=\linewidth]{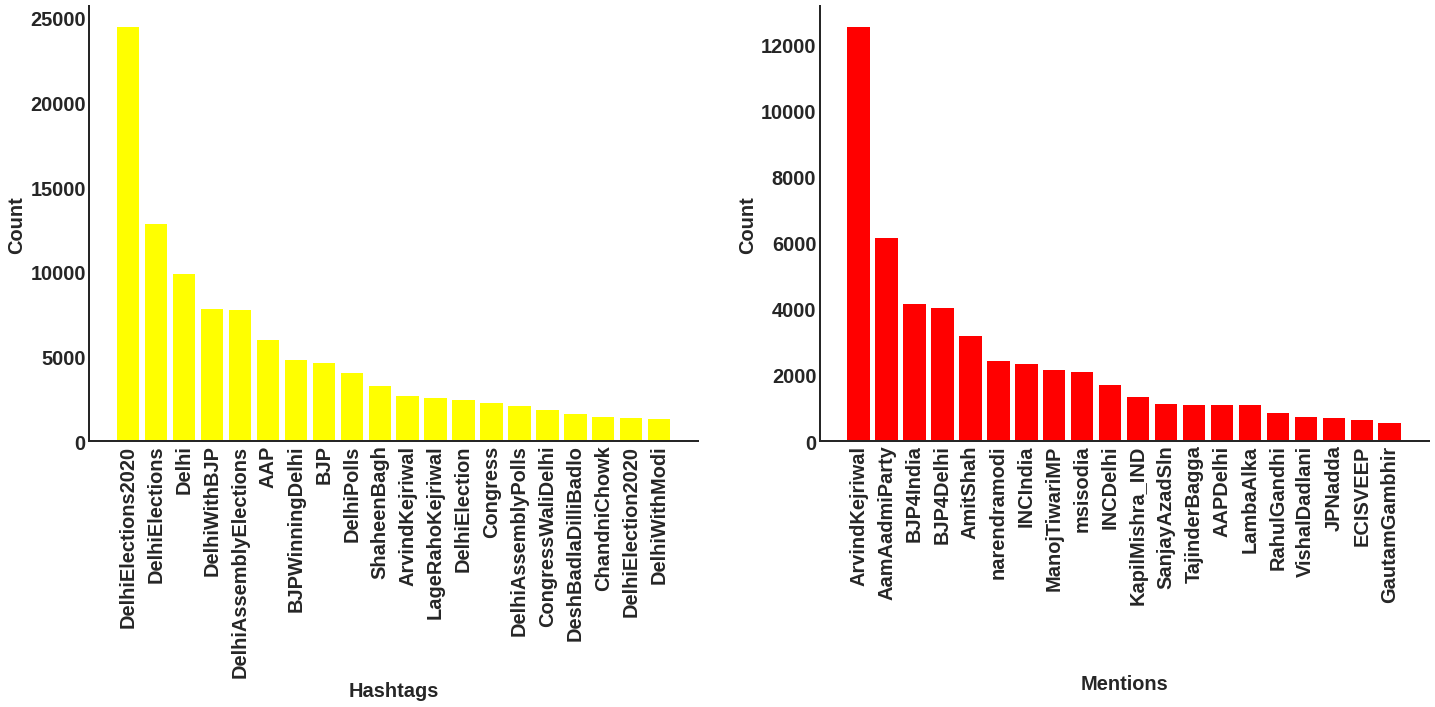}
  \caption{(a) Top 20 hashtags and (b) Top 20 mentions, based on the number of tweets containing those hashtags or mentions.}
  
  \label{fig:hashment}
\end{figure*}

The use of hashtags or mentions are quite common on Twitter for escalating the outreach of any given tweet \cite{enli2018social,shao2018spread}.
%The use of hashtags is quite common on Twitter, which in our case could be the name of a political party, candidates, star campaigners of the election or related to any issue relevant to the election. Similarly, Twitter mentions, i.e., Twitter handles in the tweets are also used to increase the reach of the tweets by mentioning Twitter accounts of users with large followers on the platform \cite{shao2018spread}.
Therefore, the top twenty hashtags and mentions, based on their usage frequency in our dataset are extracted and shown in Fig.~\ref{fig:hashment}. We observe that the top hashtags in the collected tweets contains all kind of hashtags, neutral as well as specific to any party or party members. The most frequently occurring hashtag specific to a party is \textit{DelhiWithBJP}, which does not coincide with the actual results of the election. In the case of mentions, the two most frequently used Twitter handles were \textit{ArvindKejriwal} and \textit{AamAadmiParty}, and they both belong to the winning party. 
%There are all kinds of hashtags present, be it neutral or specific to any party or party members.

\subsubsection{Keywords based Analysis}
Since mentioning a party need not necessarily be done through symbols '\#' or '@'; hence a keyword-based approach is used with the help of names and abbreviations of all three political parties. There were total 91,700 tweets that stated the name of any political party; hence a comparison of the number of tweets mentioning the parties is shown in Fig.~\ref{fig:mentions}(a). The party with the maximum number of tweets mentioning its name is BJP (Bharatiya Janta Party), whereas the party that won the election is AAP (Aam Aadmi Party).

\subsubsection{Using Tweets Mentioning only one Party}

\begin{figure*}
  \centering
  \includegraphics[width=\linewidth]{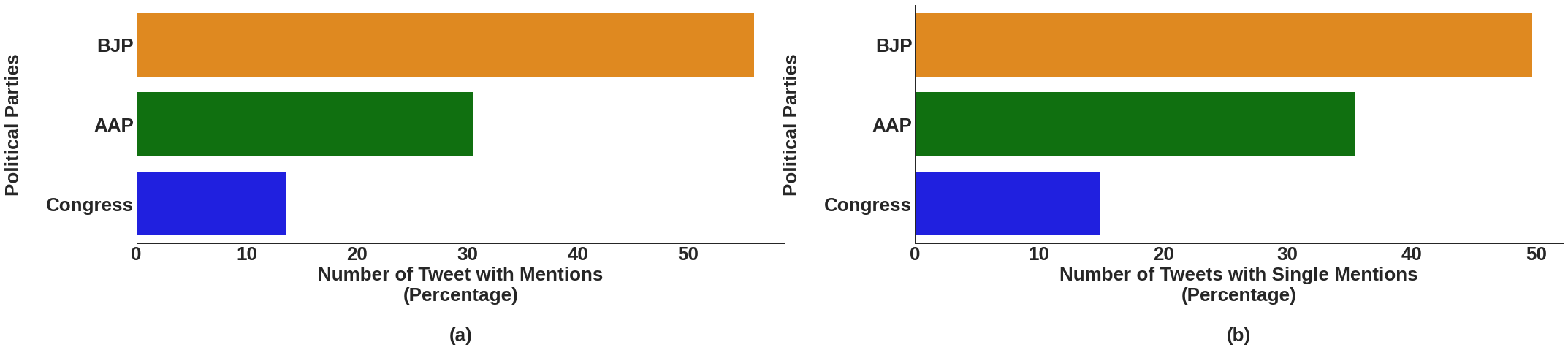}
  \caption{Share of mentions among the political parties using (a) tweets mentioning at least one party and (b) tweets mentioning only one party.}
  \label{fig:mentions}
\end{figure*}

As one tweet may refer to more than one party, therefore we extracted 62,978 tweets mentioning only single party \cite{sang2012predicting} to get a more clear picture in terms of party mentions. The comparison between the three political parties is shown in Fig.~\ref{fig:mentions}(b). Here, it was again observed that BJP was getting more mentions than the winning party. However, this analysis does not consider the sentiments of the tweets to verify if the mentions are in the support of the party or not; we will discuss this later.
%However, this analysis does not consider that the mentions were in support of the party or not; we will discuss this later.

\subsection{Analyzing Sentiments towards Party}

\begin{figure}
  \centering
  \includegraphics[height=6cm,width=8cm]{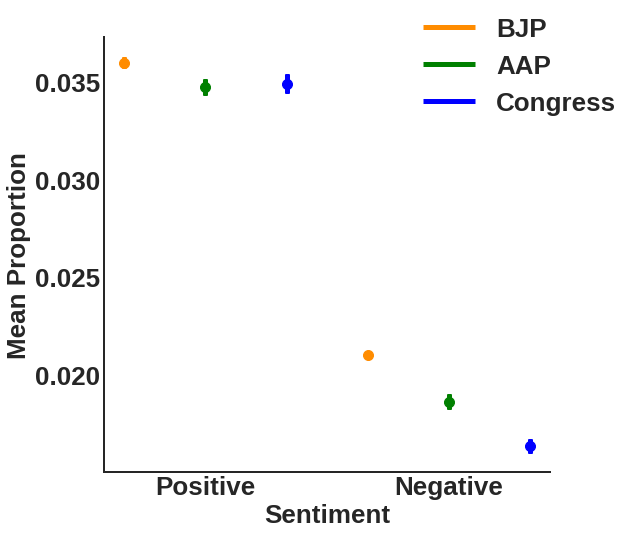}
  \caption{Mean proportion of positive and negative sentiment for all three political parties using tweets mentioning only one party.}
  
  \label{fig:psent}
\end{figure}

The number of tweets mentioning any party just gives an indication of the most talked about party on Twitter and may not highlight the fact that whether the discussions were in favor of the party or not \cite{sang2012predicting}. Hence, it is vital to study the sentiments towards those parties, as done in previous works \cite{sang2012predicting,bermingham2011using,sharma2016prediction,singh2020can,srivastava2015analyzing}. The tweets mentioning a single party were used for analyzing sentiments using lexicon based approach, which is explained in detail in Section \ref{sec:sent}. The mean proportion of both positive and negative emotions are shown for all three political parties in Fig.~\ref{fig:psent}. The figure shows that both positive and negative sentiments are more for BJP as compared to other parties.
%The mean proportion of both positive as well as negative emotions are compared in the tweets belonging to the entire period of data collection and is shown for all three political parties in Fig.~\ref{fig:psent}. As seen from the figure, both positive and negative sentiments are more for BJP as compared to other parties.

\begin{figure*}
  \centering
  \includegraphics[width=\linewidth]{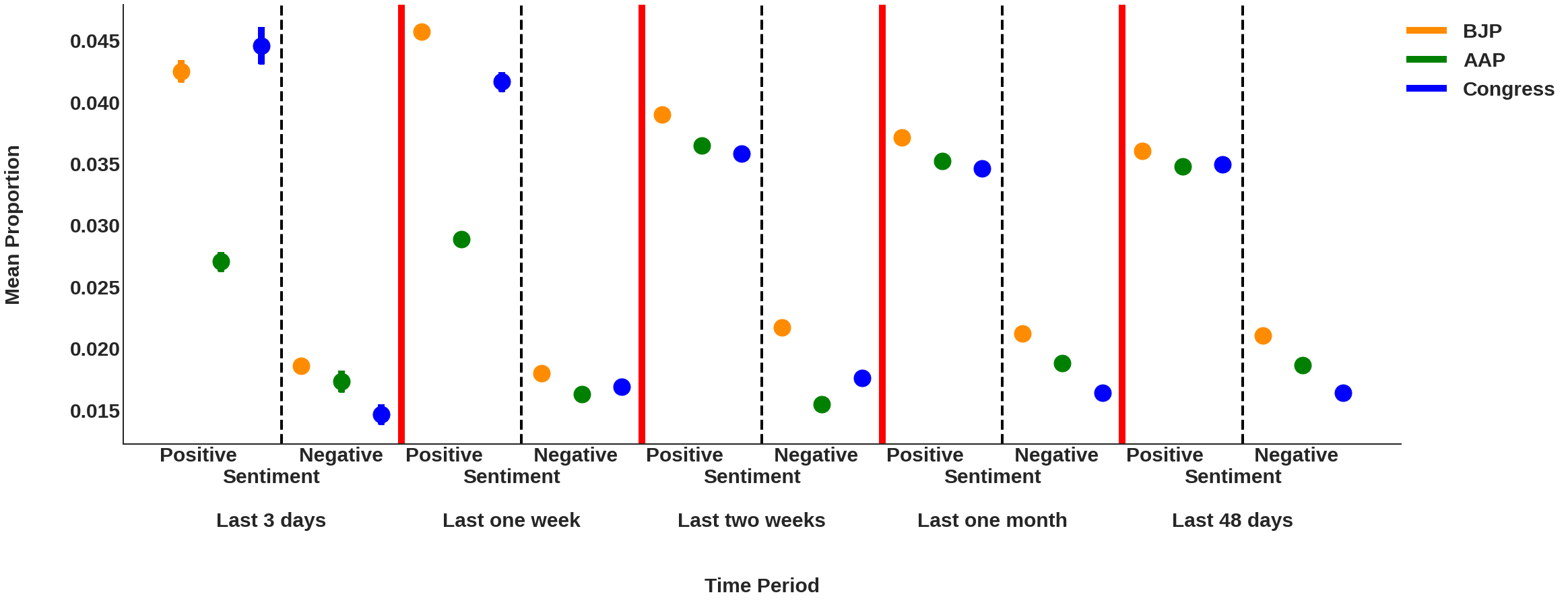}
  \caption{Comparing mean proportion of positive and negative sentiments towards political parties using different time periods.}
  
  \label{fig:ptsent}
\end{figure*}

In the above analysis, the complete data of 48 days window was used for comparing sentiments towards each party. Assuming that the time difference from the election might affect the resultant sentiments, we perform a time-based analysis of sentiments towards political parties (Fig.~\ref{fig:ptsent}). As seen from the figure, the proportion of positive sentiments is mainly dominated towards BJP for all different window sizes, except for the last three days, where Congress was leading in the positive sentiment. The winning party AAP does not lead in any window size for positive sentiments. In the case of negative sentiments, if we consider only the last one week, AAP had the lowest average proportion with respect to BJP (\textit{Effect-Size} = $0.46$ and \textit{p-value} $\leq$ 0.0001) and Congress (\textit{Effect-Size} =  $0.47$ and \textit{p-value} $\leq$ 0.0001), where Effect size and p values are computed using Mann Whitney U Test \cite{mann1947test}. Similarly, for the last two weeks, AAP again had the least proportion of negative sentiments as compared to both BJP (\textit{Effect-Size} = $0.43$ and \textit{p-value} $\leq$ 0.0001) and Congress (\textit{Effect-Size} = $0.48$ and \textit{p-value} $\leq$ 0.0001). These observations conflict with previous studies that mainly rely on sentiments (positives, negatives, and both) for predicting elections \cite{singh2020can,bermingham2011using}. Our results show that the lower negative sentiments as compared to higher positive sentiments are a better indication of the winning.
%\footnote{ES is the effect size, and p is the probability in Mann Whitney U Test; please refer \cite{mann1947test} for further details.}
%In the above analysis, a 48 days window was used for comparing sentiments towards each party. Assuming, that a window size might play any role in the resultant sentiments, therefore a time-based analysis of sentiments towards political parties was also done (Fig.~\ref{fig:ptsent}). As seen from the figure, the proportion of positive sentiments is mainly dominated by BJP for all different window sizes, except for the last 3 days, where Congress was leading in the positive sentiment. But the window size of 3 days is too short to consider it as concrete evidence and important point to be considered is in terms of positive sentiments, the winning party i.e. AAP does not lead in any window size. In case of negative sentiments, if we consider only last one week,AAP was having lowest average proportion with significantly different distribution \cite{mann1947test} with respect to BJP (ES = 0.46 p $\leq$ 0.0001) and Congress (ES = 0.47 and p $\leq$ 0.0001). Similarly, for last two weeks, again AAP had the least proportion of negative sentiment with significant distribution with respect to both BJP (ES = 0.43 and p $\leq$ 0.0001) and Congress (ES = 0.48 and p $\leq$ 0.0001). These observations conflict with previous studies that mainly relies on sentiments for predicting elections \cite{singh2020can,bermingham2011using}.

\subsection{Candidate's Availability on Twitter}

\begin{figure}
  \centering
  \includegraphics[width=.6\linewidth]{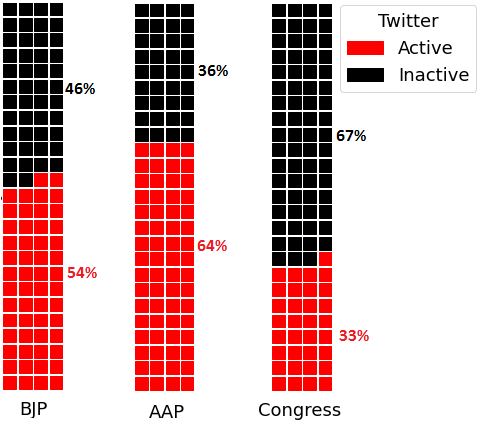}
  \caption{Availability of candidates on Twitter for all three political parties.}
  
  \label{fig:cpart}
\end{figure}

All candidates from three political parties were not available on Twitter; hence it becomes vital to compare the level of participation at the party level. A candidate is considered available if there is at least one tweet from the candidate during our collection period. The participation of the candidates from each party is shown in Fig.~\ref{fig:cpart}. The winning party has the maximum share of candidates available on Twitter, followed by BJP and Congress. This could be one of the reasons that might have made AAP candidates popular among the voters. Next, we compare the activeness of different participants on Twitter. In our analysis, the activeness refers to the number of tweets by the candidates during our data collection period. The activeness is studied from two perspectives, first on a daily basis and second using multiple time frames, each spanning over three days.
%Even though, there is unequal participation by the candidates of different political parties, another important factor that needs to be assessed is to compare their activeness on the platform. 

\subsubsection{Daily Tweeting Activeness}

\begin{figure}
  \centering
  \includegraphics[width=8cm]{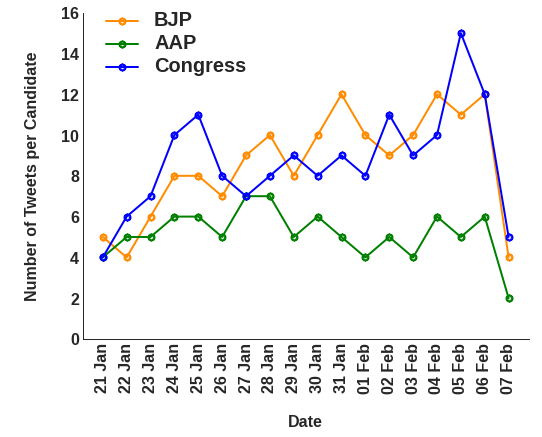}
  \caption{Daily basis comparison of the number of tweets per candidate of the three political parties.}
  
  \label{fig:daily}
\end{figure}

Here we compare the tweeting activity of the candidates of all the parties on a daily basis for a period of 18 days (the pre-election period when all the parties have declared their candidates for every seat) before the election day. Considering the fact that the number of candidates of each party on Twitter is not the same, therefore instead of comparing the total number of tweets by all the candidates of each party, we compare the normalized value, i.e., the average number of tweets per active candidate (Fig.~\ref{fig:daily}). The overall activeness per candidate is somewhat similar for BJP and Congress, where either of them leads on any particular day. One important observation is that although AAP has more candidates on Twitter, they are relatively less active for the entire duration. This may indicate that more activeness prior to election does not ensure a win for any party. One interesting observation is that the overall activeness was low for all the parties on the last day before the election. The passiveness on the last day is due to the `pre-election silence', as a political party is not allowed for promotional activities from a day before the election date. This pre-election silence period varies among different countries.
%We may compare passiveness on the last day to the 'pre-election silence', where a political party is not allowed for promotional activities in several countries, a day before the election date \cite{wikipedia_2020}. 

\subsubsection{Time Frame Analysis of Activeness}

\begin{figure*}
  \centering
  \includegraphics[width=.9\linewidth]{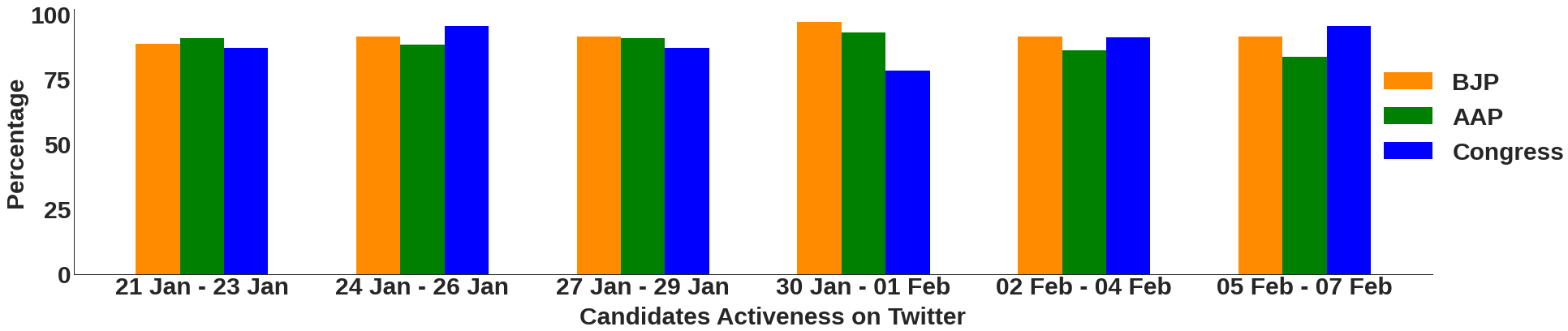}
  \caption{Activeness of candidates of different political parties on each window of size 3 days, spanned over the 18 days of data collection.}
  
  \label{fig:time_part}
\end{figure*}

The level of participation may vary from one candidate to another on a daily basis. Hence, we used a window of three days as one time frame to average out the frequent changes in the participation of the candidates. As stated earlier, by participation, we mean the candidate must have tweeted at least once in the given time frame. The time frame comparison of candidates' participation on Twitter is shown in Fig.~\ref{fig:time_part}. The participation by the candidates available on Twitter does not vary much for all the different time frames. There was slight variation for the AAP party; it gets less involved in tweeting as we move closer to the election. On the other hand, Congress was relatively less involved during the mid-interval and this might be due to the major political campaigns in that period by other parties i.e AAP and BJP \cite{jagranenglish_2020}. 
%of our collection period.

\subsection{Comparing Twitter Profiles}

Although an election might be conducted on a single day or a span of few days or even in months, a candidate or a party may run their political campaign for a longer period to widen their reach among voters. The comparison of various attributes of Twitter profile (Refer Section \ref{sec:profile} for identifying profile details) is performed from the following two perspectives.

\subsubsection{Candidates}
\begin{figure*}
  \centering
  \includegraphics[width=.95\linewidth]{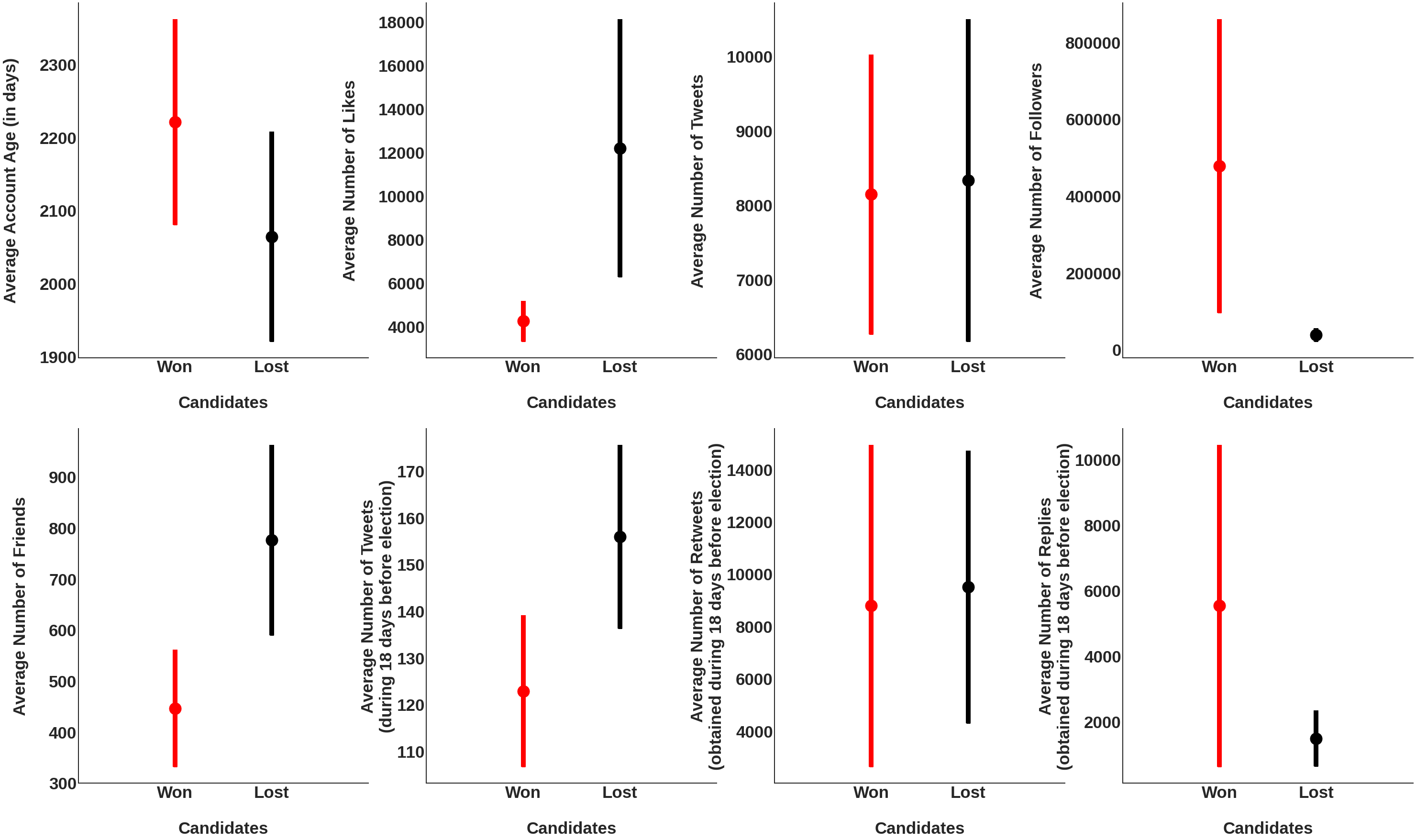}
  \caption{Comparing Twitter Profiles of the winning and loosing candidates.}
  
  \label{fig:cprofile}
\end{figure*}

The Twitter profile of the winning candidates is compared with the losing candidates using the mean of all the attributes, as shown in Fig.~\ref{fig:cprofile}. The average number of followers is higher for the candidates who won the election than those who lost the election (\textit{Effect-Size} = $0.32$ and \textit{p-value} $\leq$ 0.0001). Thus, having a large number of followers on Twitter might highlight a candidate's popularity among the voters \cite{cameron2016can}. In the case of other attributes, we did not observe any significant differences, which conflicts with the previous study that associates candidate's twitter usage \cite{bright2020does} and the response to a candidate with election outcomes \cite{digrazia2013more}.%the amount of Twitter usage with vote shares \cite{bright2020does}.  

\begin{figure*}
  \centering
  \includegraphics[width=\linewidth]{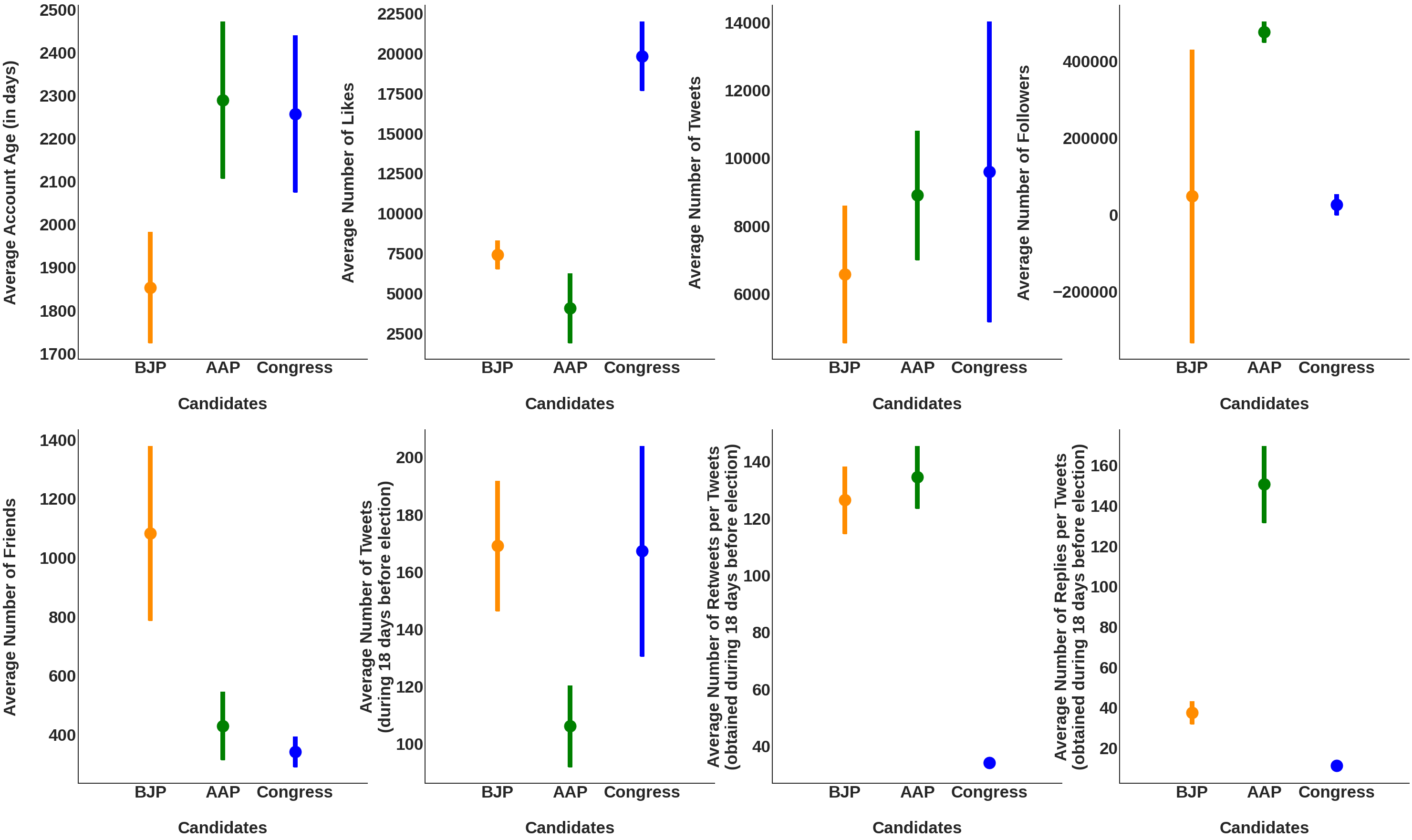}
  \caption{Comparing Twitter Profiles of the candidates of all three political parties.}
  
  \label{fig:cpprofile}
\end{figure*}

We also compared the Twitter profiles of candidates with respect to their political parties, as shown in Fig.~\ref{fig:cpprofile}. The average number of followers of all the candidates of the winning party (i.e., AAP) was higher, with their distribution being statistically significant, as compared to BJP (\textit{Effect-Size} = $0.30$ and \textit{p-value} $\leq$ 0.01) and Congress (\textit{Effect-Size} = $0.30$ and \textit{p-value} $\leq$ 0.01). Similar was the case with the average number of replies per tweet with the AAP party having a value greater than BJP (\textit{Effect-Size} = $0.39$ and \textit{p-value} $\leq$ 0.0001) and Congress (\textit{Effect-Size} = $0.48$ and \textit{p-value} $\leq$ 0.05). If we compare these findings with the ones depicted in Figure \ref{fig:cprofile}, the absence of significant results in the latter in terms of response to a candidate, might indicate the considerable differences in the popularity of the candidates belonging to the winning party i.e. AAP.%, in terms of number of followers, the canidates of AAP seems to be popular with respect to other candidates. On the contrary, in terms of response to a candidate, 
.%Thus the popularity of a candidate on Twitter in terms of the number of followers matters in winning the elections, as shown in a previous study also \cite{cameron2016can}.   

\subsubsection{Political Parties}
\begin{figure*}
  \centering
  \includegraphics[width=\linewidth]{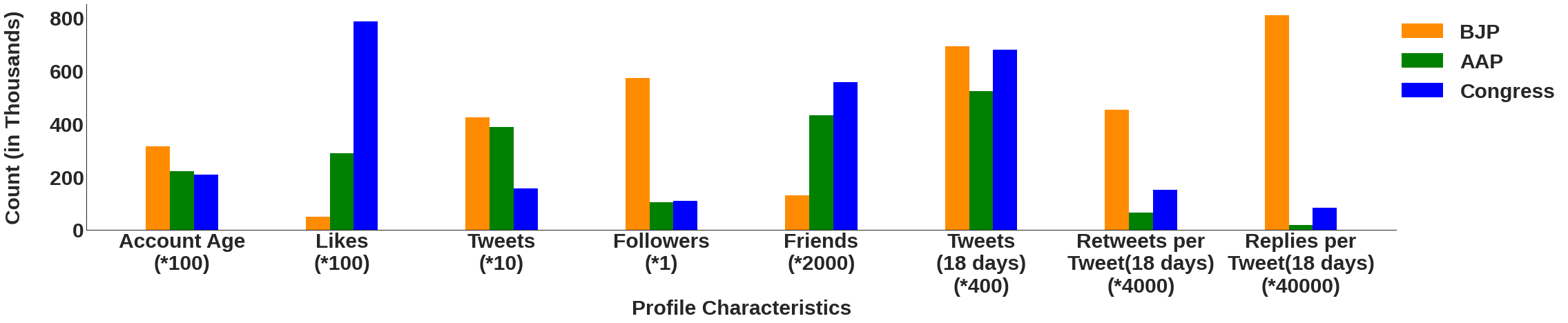}
  \caption{Comparing Twitter Profiles of the political parties using their Delhi specific Twitter handles. All the attributes were scaled for displaying on one plot, where the scaling factor for each attribute is defined below the axis label itself.}
  
  \label{fig:pprofile}
\end{figure*}

The Twitter profiles of three major political parties are also compared, and the results are shown in Fig.~\ref{fig:pprofile}. Here, for the sake of simplicity, all the attributes are shown together with the help of scaling, as we are only interested in relative values and not absolute values. The winning party (i.e., AAP) does not lead in any of the attributes. On the other hand, BJP has a very good response to their tweets in terms of retweets as well replies and also has a relatively large number of followers. This could be due to it being the largest party in the country in terms of the number of its representatives at the national as well as states level \cite{wikipedia_2020a}. Hence the Twitter profiles of political parties may not be a good indicator of election results.      

\subsection{Predicting Candidates' Win or Lose}\label{mlmodel}

The tweets of the candidates were used to develop a machine learning model that can guide us in predicting the probability of winning the election for a candidate. The details of model development are given in Section \ref{sec:model}. The results of assessing the performance of different machine learning models on various combinations of word embedding, moral, and linguistic features are shown in Table \ref{tab:single}.  The performance of Random Forest Classifier was better than other machine learning models, based on the F1 score. If we compare the scores for each kind of feature, then the simple bag of words model performed relatively better. The linguistic features resulted in providing an F1 Score of around 80 percent, thus indicating the power of the lexicon based method to differentiate the tweets from the winners of the election. The five basic moral dimensions that we considered exhibited poor performance. We also combined the best performing bag of words features with linguistics and moral dimensions and trained the models as shown in Table \ref{tab:combined}. We observe that the bag of words features provide best results even when compared with a combination of different features. These results are promising and will motivate the scientific community that the ML models trained on the features extracted from the tweets of the candidates can guide us in predicting the elections' outcome. This novel approach can help in designing better election prediction models using social media data. In future, we will work on designing the prediction model using the datasets from different elections from countries across the globe, and will analyze the performance on different datasets for a better understanding. %The linguistic features were also able to give an F1 Score of around 80 percent, thus indicating the power of the lexicon based method to differentiate the tweets from the winners of the election.%This experiment is out of scope for the current work.
%The results obtained using only the bag of words features could not be improved through several combinations. Such kind of study is a motivation to the researchers, that features extracted from the tweets of the candidates coupled with a machine learning model can guide us in predicting the elections. Our analysis opens a broad area of research, where merging several different election data together from countries across the globe and extracting features from the candidates' tweets might be a good approach in election predictions from Twitter.

%\begin{comment}
\begin{table}[b]
 
    \centering
    %\captionsetup{width=1.1\textwidth}
    \caption{F1 Score using various machine learning models on each given features. The models used are Support Vector Machines with Radial Basis Function (SVM), Logistic Regression (LR), Decision Tree (DT), Random Forest (RF), Adaboost (AB) and Gaussian Naive Bayes (GNB).}
    \label{tab:single}
    \begin{tabular}{|c|c|c|c|c|c|c|}
    \hline
    \multirow{2}{*}{Features} &  \multicolumn{6}{c|}{Machine Learning Models}\\
    \cline{2-7}
    & SVM & LR & DT & RF & AB & GNB\\
    \hline
    Bag of Words & 86.98 & 85.10 & 83.94 & \textbf{87.72} & 79.17 & 80.15\\
    \hline
    TF-IDF & 87.11 & 84.37 & 84.68 & \textbf{87.44} & 78.55 & 80.55\\
    \hline
    Word2Vec & 80.92 & 70.30 & 76.12 & \textbf{83.99} & 69.13 & 65.18\\
    \hline
    Linguistic & 71.03 & 61.88 & 75.18 & \textbf{80.22} & 61.59 & 51.97\\
    \hline
    Moral & 54.90 & 52.66 & 61.07 & \textbf{62.31} & 55.41 & 51.33\\
    \hline

    \end{tabular}
\end{table}
%\end{comment}

\begin{table}[t]
    \centering
    %\captionsetup{width=1.1\textwidth}
    \caption{F1 Score using combinations of features on different machine learning models.}
    \label{tab:combined}
    \begin{tabular}{|p{2cm}|c|c|c|c|c|c|}
    \hline
    \multirow{2}{*}{Features} &  \multicolumn{6}{c|}{Machine Learning Models}\\
    \cline{2-7}
    & SVM & LR & DT & RF & AB & GNB\\
    \hline
    Bag of Words $+$ Linguistic & \textbf{87.28} & 85.38 & 81.21 & 86.01 & 78.74 & 81.66\\
    \hline
     Bag of Words $+$ Moral &\textbf{86.81} & 85.80 & 81.83 & 86.63 & 78.23 & 81.48\\
    \hline
    Linguistic $+$ Moral & 70.50 & 62.67 & 74.45 & \textbf{81.29} & 63.46 & 54.27\\
    \hline
    Bag of Words $+$ Linguistic $+$ Moral & \textbf{87.54} & 85.22 & 79.78 & 85.43 & 78.23 & 81.07\\
    \hline

    \end{tabular}
\end{table}

\section{Methodology}

\subsection{Tweets Sentiment Extraction}\label{sec:sent}

All election relevant tweets containing only single party mentions were used for extracting sentiments with the help of the LIWC (Linguistic Inquiry and Word Count) tool \cite{tausczik2010psychological}. LIWC tool uses a dictionary for extracting various linguistic features. Since we are interested in identifying sentiments in the tweets, we only extract word counts related to positive and negative emotions for each tweet. These word counts were normalized by dividing with the total word count of a tweet, given as one of the features in the tool. Using this, we were able to obtain the proportion of both positive as well as negative emotions in all the tweets.

\subsection{Extracting Twitter Profile Attributes}\label{sec:profile}

The various attributes of candidates and political parties profiles were extracted from Twitter and are divided into two categories.

\subsubsection{Long term attributes}
The long-term attributes are not specific to the pre-election period but rather related to the overall profile on Twitter. These include account age, number of tweets, likes, followers, and friends on Twitter. All these attributes were collected using the Twitter REST API.

\subsubsection{Short term attributes}
The short-term attributes include number of tweets by an account, the average number of retweets and replies. Here short term corresponds to the period of our data collection.% (for candidate related as well as party related tweets). %are specific to our data collection period, such as the number of tweets by an account, the average number of retweets and replies received by the account per tweet created within 18 days before the election. These attributes were directly obtained from our candidate related tweets dataset and party related dataset.

\subsection{Model Development}\label{sec:model}

\begin{figure}
  \centering
  \includegraphics[width=\linewidth]{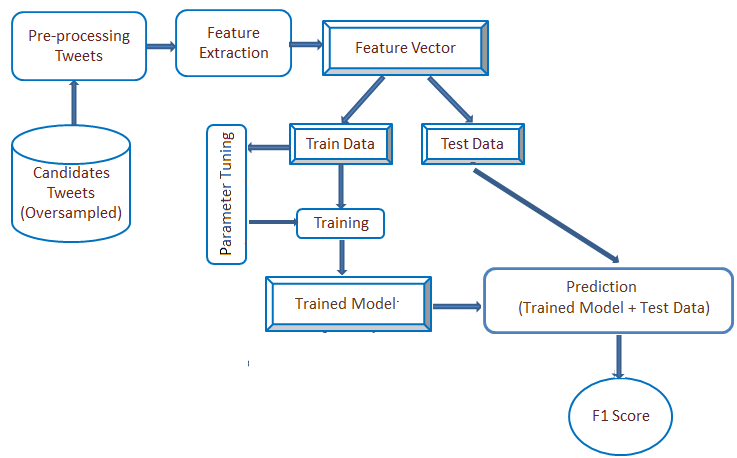}
  \caption{Different phases involved in learning and predicting tweets by the winning and loosing candidates.}
  
  \label{fig:model}
\end{figure}

In this section, we discuss the steps involved in developing a binary classifier for predicting whether a tweet is by winning or losing candidates (Fig.~\ref{fig:model}). 

\subsubsection{Sampling}
The tweets of all the candidates were used as sample data. The labels for the samples are based on the tweet's author, i.e., if the tweet is by a winning candidate, then we consider it as a positive sample, and if it is by a non-winning candidate, then we label it as a negative sample. Thus, we had a total of 5282 positive samples and 9041 negative samples. Since the frequency of negative samples is almost double than their counterpart, therefore we employed an oversampling technique \cite{chawla2004special} in order to balance the data, and finally, there were 18082 samples in our dataset.

\subsubsection{Pre-processing}
The commonly used steps for pre-processing tweets were applied, such as convert to lower case, remove URLs, punctuations, extra spaces, non-ASCII characters, hashtag symbol, mention symbol, and stopwords.

\subsubsection{Feature Extraction}

%As the sampled data was in the form of a text, hence it was required to convert the same into a feature vector. The following types of features were extracted from the data to be tested separately.
We use the following features from the tweet text to generate feature vector.

\paragraph{Bag of Words}

The frequently occurring 5000 words in the sampled data and their frequency were used as the features.

\paragraph{TF-IDF}
%Knowing the fact that a given word may have certain importance with respect to a particular tweet and overall tweets, therefore, the best 5000 features were retained using the TF-IDF approach applied for every word.  
A given word may have certain importance with respect to a particular tweet and overall tweets, therefore, the best 5000 words were retained using the TF-IDF approach.

\paragraph{Word2Vec Features}
The glove model, i.e., trained on a Twitter dataset \cite{pennington2014glove}, was used to convert each word in the tweet into a 200-dimensional vector. Next, the vectors for all words were averaged to form a single vector of 200 dimension for each tweet that can be used as features.
%The glove pre-trained model on a Twitter dataset having the dimension of 200 \cite{pennington2014glove}, was used to convert each word in the tweet into a vector and later averaged out to form a single vector of size 200 for each tweet.

\paragraph{Linguistics}
The linguistic features of all the tweets, as obtained from the LIWC 2015 dictionary \cite{tausczik2010psychological} were used. The tool only provides the word count, therefore, we converted those word counts into the normalized proportion of each feature using the word count of each tweet. 

\paragraph{Moral Values}
The moralstrength tool, developed by Araque et al. \cite{araque2020moralstrength}, was used to predict moral foundations for each sample. The given tool estimates the proportion of five basic moral foundations. 
%The given tool uses a dictionary to estimate the proportion of five basic moral foundations. 

%\subsubsection{Data Partitioning}
%The feature vector obtained from each of the above categories in the previous step, were partitioned into train and test set with a ratio of 0.7 : 0.3.

\subsubsection{Training and Testing set}
The feature vector obtained from each of the above categories in the previous step, were partitioned into train and test set with a ratio of 0.7 : 0.3. We use six machine learning models namely Support Vector Machines (SVM), Logistic Regression (LR), Decision Tree (DT), Random Forest Classifier (RFC), AdaBoost (AB), and Gaussian Naive Bayes (GNB) \cite{bonaccorso2017machine}. The models are trained and tested for different category of features; more details are available in Section \ref{mlmodel}.
%Using the model parameters obtained from the previous step, all category of features were separately trained and tested on all six machine learning models. 

\subsubsection{Parameter Tuning}
The setting of parameter values to achieve optimal result is an important component in the design of any machine learning model. In this regard, the train set was used to tune the model parameters using 5-fold cross-validation. 

\subsubsection{Evaluation Criteria}
As F1 score evaluates both precision as well as recall capability of a machine learning model, hence we have used F1 score for the assessing the performance of all the models.%As the data was explicitly made balanced using an oversampling technique, hence instead of using accuracy, 
%we have used f1 score as an evaluation criterion. The average f1 score was computed for both cross validation as well as the test set.

\section{Conclusion}
The use of online social networks by political parties for propagating their agendas and ideologies is getting very common. In this study, we use a bi-level approach to analyze the correlation of Twitter activities with the election winners both at the party level and candidate level. We observed that the widely used measures such as party mentions and sentiment towards the party were unable to identify the winning party. In the case of Twitter profiles of the candidates, the winning candidate has relatively more followers and a higher number of the average replies per tweet. Next, we train machine learning models on the generated features using Twitter election data to predict whether the tweet is from the winning candidate or not.  We observe that the Random Forest model gives promising results that can be used further to predict the winning party. Such models, if trained on different election-related data, may open new directions of research that can aim at predicting the probability of winning for a candidate. In future work, one can use more parameters for training the model, such as network-based features of users, re-tweet network with labels (positive/negative/neutral), the stance of users, etc. In this work, we have used the Twitter data for one election; however, in the future, researchers can assess the capabilities of Twitter by performing a cross-national study, covering countries with different economic and demographic backgrounds.

\bibliographystyle{unsrtnat}
\bibliography{delhi}  %%% Uncomment this line 

\end{document}